# Efficient Quality-Based Playout Buffer Algorithm

Emine Zerrin Şakir
Mango d.o.o
Valvasorjeva ul. 12a
2000 Maribor, Slovenia
emine.sakir@mangodoo.net

Christian Feldbauer
Graz University of Technology
SPSC Lab.
8010 Graz, Austria
feldbauer@tugraz.at

## Abstract

*Playout buffers are used in VoIP systems to compensate for network delay jitter by making a trade-off between delay and loss. In this work we propose a playout buffer algorithm that makes the trade-off based on maximization of conversational speech quality, aiming to keep the computational complexity lowest possible. We model the network delay using a Pareto distribution and show that it is a good compromise between providing an appropriate fit to the network delay characteristics and yielding a low arithmetical complexity. We use the ITU-T E-Model as the quality model and simplify its delay impairment function. The proposed playout buffer algorithm finds the optimum playout delay using a closed-form solution that minimizes the sum of the simplified delay impairment factor and the loss-dependent equipment impairment factor of the E-model. The simulation results show that our proposed algorithm outperforms existing state-of-the-art algorithms with a reduced complexity for a quality-based algorithm.*

## 1. Introduction

Quality of Service (QoS) is an important metric for VoIP. Jitter, delay, and loss are the main network parameters that affect the perceived speech quality. Playout buffers can compensate for jitter and resynchronize the received packets. Since playout buffers do not play the packets back as soon as they are received, but wait for a certain time in order to play them back in a continuous way, the overall delay increases. On the other hand, packets arriving after their playout time are lost, which increases the total packet loss rate. For those reasons, the playout buffer size is a trade-off between delay and loss.

Different playout buffer algorithms have been developed since the early 1980s. Most of these algorithms do not even try to make the trade-off between delay and loss, but rather set a threshold to the playout delay or to the loss. These approaches fail in case of unexpected network conditions because of under- or overestimating the playout delay. As it is the perceived speech quality that is important for VoIP applications, it is better to develop a playout buffer algorithm that maximizes the conversational speech quality.

Figure 1 shows the basic approach of quality-based playout buffer algorithms [5]. In this figure, $d$ refers to the delays the packets experience in the network whereas $P_D$ refers to the end-to-end delay (playout delay). Most of the quality-based algorithms find the playout delay by means of a search algorithm [7, 13, 4, 6] that maximizes the quality factor or minimizes the impairments. Such a search results in a high computational complexity.

Our goal in this paper is to find a playout buffer algorithm that is quality-based and gives the optimum playout delay by means of a closed-form solution with low arithmetical complexity. To achieve this, we modify the E-Model [8] and simplify its delay impairment calculation. We model the delay distribution using a Pareto distribution, which fits well to the network characteristics and is easy to compute. Our playout delay is the one, that minimizes the sum of the simplified delay impairment and the loss-dependent equipment impairment factor of the E-Model without any iterative or search methods. We also take into account the burstiness of the loss pattern in our calculations. Except in [6], in all other works, burstiness was ignored, and loss was supposed to occur randomly. The proposed quality-based playout approach is shown in Figure 2.

Our proposed algorithm is introduced in the second section of this paper. In the third section, the simulations and the results are shown. Section four discusses the developed method, and section five concludes the paper.

## 2. Proposed playout buffer algorithm

As we want to develop a playout buffer algorithm based on maximizing the conversational speech quality, we require a quality model and choose the E-Model. The output

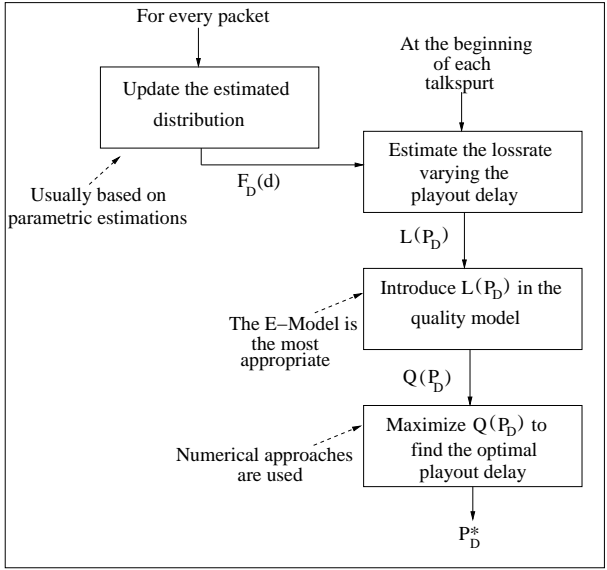

**Figure 1. Quality-based playout approach.**

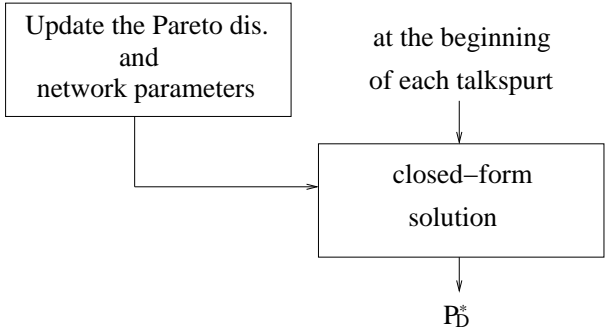

**Figure 2. Proposed quality-based playout approach.**

of the E-Model is a single scalar $R$ factor which is defined as

$$R = R_0 - I_{e-eff} - I_{dd}, \quad (1)$$

where $R_0$ is the basic signal-to-noise ratio, $I_{e-eff}$ is the loss-dependent equipment impairment factor, and $I_{dd}$ is the delay impairment factor caused by a too long overall end-to-end delay (or playout delay $P_D$). The calculation of the delay impairment factor as defined in the E-Model [8] is rather complex. As we want to reduce computational complexity and are aiming for a closed-form solution to find the optimum playout delay, we simplify the delay impairment factor as

$$I_{dd} = \begin{cases} 0 & \text{for } P_D < 150\text{ms} \\ \log\left(\frac{P_D}{150}\right) \cdot 55 & \text{for } P_D \geq 150\text{ms}. \end{cases} \quad (2)$$

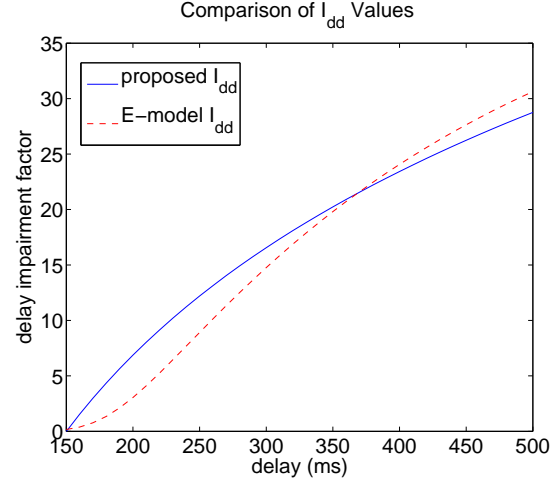

**Figure 3. Comparison of $I_{dd}$ values.**

Figure 3 shows the comparison of the delay impairment factor of the E-Model and our proposed simplified version. It can be seen that our calculation is close to the calculation of the E-Model, but analytically much easier tractable. This will be of importance in section 2.2.

The loss-dependent equipment impairment factor as defined in the E-Model is

$$I_{e-eff} = I_e + (95 - I_e) \cdot \frac{L}{\frac{L}{BurstR} + B_{pl}}, \quad (3)$$

where $I_e$ and $B_{pl}$ are codec-specific values, which can be found in ITU-T Recommendation G.113 [9]. $BurstR$ is the burst ratio and can be calculated as in eqn. (4) for packet loss processes modeled by a 2-state Markov model (Gilbert model) with transition probabilities $p$ from received to lost states and $q$ from lost to received states [10].

$$BurstR = \frac{1}{p+q} = \frac{L/100}{p} = \frac{1 - L/100}{q} \quad (4)$$

$L$ is the overall loss rate in percent which includes both losses due to network failures and a too short playout buffer. To find the loss due to the playout buffer, we have to model the delay with an appropriate distribution, as the loss due to buffering ($\rho_b$) can be described with the cumulative distribution $F(\cdot)$ of the delays the packets experience because of the network travel ($d$).

$$\rho_b = P(d \geq P_D) = 1 - F(P_D). \quad (5)$$

As seen in eqn.n(5), the loss due to buffering leads to the complementary cumulative distribution function (CCDF).

## 2.1. Modeling the delay distribution

To find an appropriate distribution, which fits well to the network delay characteristics and yields a low arithmetical complexity, we first collected necassary data by using a UDP/IP probe tool [1]. It consists of client/server programs which run on a local host and a remote host. UDP/IP packets are sent and echoed back between these two hosts. It is possible to find the delays (one or 2-way) that the packets experience in the network as well as the losses.

We have established a connection between China and Austria. The size of the probe packets was set to 64 bytes and the interval between successive packets to 30ms. We have collected data for 6 hours and taken 8 different parts of this trace data for our simulations which represent different network characteristics.

To find the best distribution, we look at the empirical cumulative distribution function (CDF) of these eight traces. The CDF of the actual data shows the characteristics of a heavy-tailed distribution. Because of that we decided to experiment our delays with Pareto, Weibull and log-normal distributions, which are heavy-tailed distributions, as well as with the exponential distribution. The CDFs of these distributions are shown in table 1.

**Table 1. CDFs of the chosen distributions.**

| Distribution | CDF ($F(d)$) |
|---|---|
| Pareto | $1 - (\frac{d_m}{d})^k$ |
| Weibull | $1 - e^{-(d/\lambda)^k}$ |
| Log-normal | $\frac{1}{2} + \frac{1}{2}\mathrm{erf}\left[\frac{\ln(d)-\mu}{\sigma\sqrt{2}}\right]$ |
| Exponential | $1 - e^{-\lambda d}$ |

Our experiments confirmed the findings of [14] that the Pareto distribution provides a good fit to the actual data, and as table 1 shows, its CDF description is less complex than for the other distributions (power function as opposed to exponential or error function). Because of these reasons we decided to use the Pareto distribution to model the delay distribution. Furthermore, we improve the quality of the fit, by fitting the Pareto distribution to the tail part only. We define the tail as all the received packets that have delays higher than the median. Under the assumptions that the optimum playout delay is practically always higher than the median and the buffer loss always lower than 50% for a VoIP application, modeling the tail is enough. The scale parameter of the Pareto distribution, which corresponds to the minimum delay value of the density, is in this case the median. The shape parameter $k$ can be found by means of maximum likelihood estimation.

Figure 4 shows the relation between the loss due to buffering and the playout delay $P_D$ where $\mu_{1/2}$ indicates the median of the delay values.

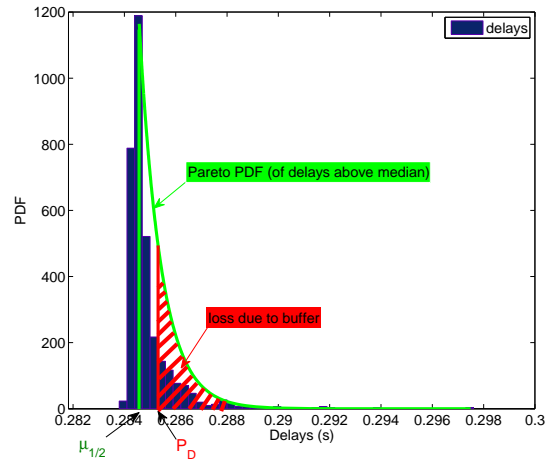

**Figure 4. Modeling the delay distribution.**

The loss rate can then be expressed as

$$L = 100 \cdot \rho_n + 100 \cdot (1 - \rho_n) \cdot 0.5 \cdot \int_{P_D}^{\infty} \mathcal{P}(\tau)d\tau \quad (6)$$

where $\rho_n$ is the loss probability due to network delivery failures. In eqn. (6), the factor 0.5 is needed because the Pareto PDF $\mathcal{P}(d)$ models only delays above the median and so, the integral must be at most 0.5 (50% of all received packets).

## 2.2. Finding the optimum playout delay analytically

Our goal is to find a playout delay which minimizes the sum of our simplified version of delay impairment factor and the loss-dependent equipment impairment factor, which are given in eqns. (2) and (3), respectively. This is equal to finding the playout delay which provides $P_D^* = \arg\max_{P_D} R$. Figure 5 shows the equipment and simplified delay impairments. As shown in this figure, $I_{dd}$ increases with increasing $P_D$, whereas $I_{e-eff}$ decreases. Therefore, the optimum playout delay $P_D^*$ is found when the derivative of $I_{dd}$ with respect to $P_D$ equals the negative derivative of $I_{e-eff}$. So the optimum playout delay is the one where

$$-\frac{d}{dP_D}I_{dd} = \frac{d}{dP_D}I_{e-eff}. \quad (7)$$

The simple form of eqn. (2) and (3) facilitates finding these derivatives, and eqn. (7) can be solved analytically. Our closed-form solution to find the optimum playout delay is:

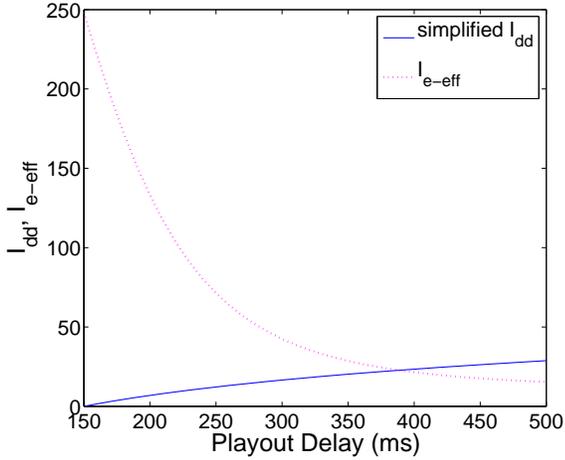

**Figure 5. Delay and equipment impairments as a function of the playout delay.**

$$\alpha_1 = k \cdot BurstR^2 \cdot (95 - I_e) \cdot \ln(10)$$
$$\alpha_2 = 110 \cdot (\rho_n \cdot 100 + BurstR \cdot Bpl)$$
$$P_D = \mu_{1/2} \left( \frac{0.5 \cdot 110 \cdot 100 \cdot (1 - \rho_n)}{\alpha_1 - \alpha_2 - \sqrt{\alpha_1(\alpha_1 - 2\alpha_2)}} \right)^{1/k} \quad (8)$$
$$P_D^* = \begin{cases} P_D, & P_D \geq 150 \\ 150, & P_D < 150 \end{cases}$$

where $\alpha_1$ and $\alpha_2$ are auxiliary variables used for a better and more readable presentation of eqn. 8.

### 2.3 Summary of the proposed algorithm

The proposed algorithm can be summarized as:

- at the beginning of each talkspurt
  - update the delay records of the last $W$ packets
  - estimate the parameters of the Pareto distribution (scale ($\mu_{1/2}$) and shape ($k$)) taking the delays above the median only
  - compute the network loss ($\rho_n$)
  - find the burst ratio using the Gilbert model
  - Find the optimum playout delay using the eqn. (8)

The value $W$ in the algorithm represents the window size.

Figure 2 demonstrates how our algorithm simplifies the quality-based approach (compare with Figure 1). We only need to estimate the parameters of the Pareto distribution, network loss and burst ratio and use them in our closed-form solution.

## 3. Simulations and performance comparison

### 3.1. Simulation set-up

We took six minutes of speech from the NTIMIT speech database. For coding we used the internet low bitrate codec (iLBC) [3] and used the voice activity detection feature of the adaptive multirate speech codec (AMR VAD option 1) [2] to find the talkspurts and silences. We did not take talkspurts shorter than 250ms as PESQ does not accept such short speech segments. We got at the end 67 talkspurts from the six minutes of speech. We used our proposed playout buffer algorithm and five other algorithms to find the playout delay. Exp-avg, f-exp-avg, min-del and spike-det are algorithms proposed by Ramjee [11], whereas p-optimum is another quality-based playout buffer algorithm [13]. In our proposed algorithm, we chose a window size of 500 packets. For the quality evaluation we used the E-Model and the conversational speech quality measurement (MOSc) [12]. The output of the E-Model is E-MOS, which is calculated by mapping the $R$ factor to MOS. MOSc is found by using the delay impairment factor of the E-Model and the loss-dependent equipment impairment factor derived from the output of PESQ. Finding MOSc is explained in detail in [12]

### 3.2. Simulation results

We found the average E-MOS and MOSc values and the total lossrate of the proposed and other algorithms using our eight traces. Our algorithm gives the best average E-MOS and MOSc values in four of these traces, the second best in three of them and the third only in one of them. Table 2 shows the results of four traces.

Figure 6 shows the average E-MOS values of all eight traces. This figure shows that the quality-based algorithms (proposed and "p-optimum") have more stable responses to different network conditions. They have reasonable behaviors at different network conditions.

This figure also shows that our proposed algorithm, compared to other algorithms, has a good performance under different network conditions.

Table 3 shows the mean E-MOS and MOSc values taken over all eight traces. Our proposed algorithm gives the best average E-MOS and MOSc values.

To compare the complexities of the quality based playout buffer algorithms, we measured the execution time of our proposed playout buffer algorithm and the p-optimum algorithm using the tic,toc function of matlab. If the number of considered playout delay values are 200, then the execution time of the p-optimum algorithm is 173.469 $\mu$s and it increases with the increasing playout delay values. The execution time of our proposed algorithm is 0.601 $\mu$s and does not have any dependency to any of the parameters as it is a closed-form solution.

**Table 2. Comparison of playout buffer algorithms.**

| Trace | Algorithm | E-MOS | MOSc | Loss (%) |
|---|---|---|---|---|
| 2 | Proposed | **3.3492** | **3.4137** | 0.8986 |
|   | p-optimum | 3.2952 | 3.3843 | 2.6122 |
|   | exp-avg | 3.3441 | 3.4022 | 0.7758 |
|   | f-exp-avg | 3.2886 | 3.3373 | 0.0614 |
|   | min-del | 3.2364 | 3.3068 | 0.5972 |
|   | spike-det | 2.4583 | 2.7810 | 15.0368 |
| 3 | Proposed | **3.7646** | **3.8220** | 0.5693 |
|   | p-optimum | 3.6910 | 3.7741 | 2.1824 |
|   | exp-avg | 3.6820 | 3.7779 | 2.7071 |
|   | f-exp-avg | 3.7611 | 3.7921 | 0.0781 |
|   | min-del | 3.6701 | 3.7341 | 2.5675 |
|   | spike-det | 2.7005 | 3.0466 | 18.3411 |
| 6 | Proposed | **3.7813** | **3.8231** | 0.5079 |
|   | p-optimum | 3.7421 | 3.7932 | 1.6075 |
|   | exp-avg | 3.7477 | 3.8001 | 1.1331 |
|   | f-exp-avg | 3.7497 | 3.7854 | 0.0391 |
|   | min-del | 3.7051 | 3.7492 | 0.4856 |
|   | spike-det | 2.9108 | 3.1840 | 21.4445 |
| 8 | Proposed | 3.9060 | 3.9464 | 0.4689 |
|   | p-optimum | 3.8438 | 3.8855 | 1.0214 |
|   | exp-avg | 3.6003 | 3.6911 | 3.0866 |
|   | f-exp-avg | **3.9220** | **3.9532** | 0.1005 |
|   | min-del | 3.6212 | 3.7283 | 3.1927 |
|   | spike-det | 3.3814 | 3.5852 | 6.3630 |

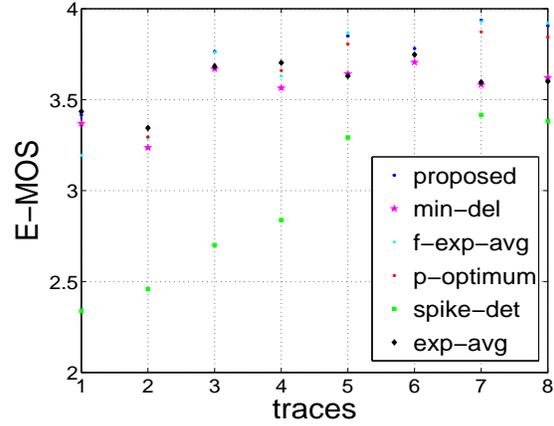

**Figure 6. Comparison of E-MOS values.**

**Table 3. Total Average E-MOS and MOSc values.**

| Algorithm | E-MOS | MOSc |
|---|---|---|
| proposed | **3.7131** | **3.7583** |
| p-optimum | 3.6677 | 3.7254 |
| exp-avg | 3.5923 | 3.6712 |
| f-exp-avg | 3.6678 | 3.7050 |
| min-del | 3.5489 | 3.6340 |
| spike-det | 2.9169 | 3.2005 |

## 4. Discussion

In our simulations we used a window size of 500 packets. But our experiments show that under different network conditions (i.e more or less stationary conditions) different window sizes are giving better results. For example, if the trace data shows stationarity, longer window sizes can be used. But in case of variable delay values, shorter window sizes should be preferred. A future work can make the window size adaptive based on the network conditions. Alternatively, a recursive estimation scheme for the parameters is possible. However also in such a scheme, the weighting factors should be adaptive with respect to the stationarity of the network conditions.

The delay impairment factor of the E-Model is known to be very strict to the delays, and its calculation is quite complex. We simplified the delay impairment factor of the E-Model, which is given in eqn. (2). Future research results for delay impairment calculations can be applied to update our method by simply changing the coefficients of eqn. (2). Furthermore, it is also possible to give control over those coefficients to the user of a VoIP application in order to adjust the buffering behavior to one's preference.

## 5. Conclusions

In this paper we propose a quality-based playout buffer algorithm, which makes the trade-off between delay and loss by maximizing the conversational speech quality. We choose to develop a quality-based playout buffer algorithm because of the fact that other methods which set a threshold to the playout delay so that the lossrate is low or negligible may overestimate the playout delay under certain network conditions and cause a dramatic decrease in conversational speech quality. Our simulation results show that the quality-based playout buffer algorithms are better adapting to different network conditions.

We try to keep the computational complexity of the algorithm as low as possible. The complexity of the quality-based playout buffer algorithms depends on the temporal resolution used in selecting the optimal playout delay. We instead find the optimal playout delay by means of a closed-form solution.

Although our proposed playout buffer algorithm does not outperform the existing state-of-the art algorithms significantly, the arguments explained in the previous paragraph makes it a better choice.